\documentclass{PoS}

\def\la{\mathrel{\hbox{\rlap{\hbox{\lower4pt\hbox{$\sim$}}}\hbox{$<$}}}}
\def\ga{\mathrel{\hbox{\rlap{\hbox{\lower4pt\hbox{$\sim$}}}\hbox{$>$}}}}

\def\deg{{^\circ}}

\def\fm{\hbox{$.\!\!^{\rm m}$}}

\def\fdg{\hbox{$.\!\!^\circ$}}

\newcommand{\kms}{{\,km\,s$^{-1}$}}

\newcommand{\HI}{\mbox{\normalsize H\thinspace\footnotesize I}}

\def\apj    {{ApJ }}

\def\mnras  {{MNRAS }}

\title{Towards a Full Census of the Obscure(d) Vela Supercluster using MeerKAT}

\ShortTitle{The VSCL MeerKAT survey}

\author{\speaker{R.C. Kraan-Korteweg}$^1$,  E.C. Elson$^1$, S.L. Blyth$^1$, C. Carignan$^1$, B.S. Frank$^1$, T.H. Jarrett$^1$, M.E. Cluver$^2$, P. Serra$^3$, G.I.G. J\'ozsa$^4$\\
              $^1$Astronomy Department, University of Cape Town,  7701 Rondebosch, South Africa\\
        \ E-mails: \email{kraan,ed,sarblyth,ccarignan,bradley,jarrett@ast.uct.ac.za}\\
       	$^2$Department of Physics, University of the Western Cape, Bellville, South Africa\\
        \ E-mail: \email{mcluver@gmail.com}\\
        $^3$Osservatorio Astronomico di Cagliari (INAF), Selargus, Italy\\
        \ Email: \email{pserra@oa-cagliari.inaf.it}\\
        $^4$SKA South Africa, Radio Astronomy Research Group, Pinelands, South Africa\\
        \ Email: \email{jozsa@ska.ac.za}
}

\abstract{Recent spectroscopic observations of a few thousand partially obscured galaxies in the Vela constellation revealed a massive overdensity on supercluster scales straddling the Galactic Equator ($\ell ~\sim 272\fdg5$) at $cz \sim 18\,000$\,\kms. It remained unrecognised because it is located just beyond the boundaries and volumes of systematic whole-sky redshift and peculiar velocity surveys -- and is obscured by the Milky Way. The structure lies close to the apex where residual bulkflows suggest considerable mass excess. The uncovered Vela Supercluster (VSCL) conforms of a confluence of merging walls, but its core remains uncharted. At the thickest foreground dust column densities ($|b| \la 6\deg$) galaxies are not visible and optical spectroscopy is not effective. This precludes a reliable estimate of the mass of VSCL, hence its effect on the cosmic flow field and the peculiar velocity of the Local Group.

Only systematic \HI-surveys can bridge that gap. We have run simulations and will present early-science observing scenarios with MeerKAT~32 (M32) to complete the census of this dynamically and cosmologically relevant supercluster. M32 has been  put forward because this pilot project will also serve as precursor project for \HI\ MeerKAT Large Survey Projects, like Fornax and Laduma. Our calculations have shown that a survey area of the fully obscured part of the supercluster, where the two walls cross and the potential core of the supercluster resides, can be achieved on reasonable time-scales (200\,hrs) with M32. 
}

\FullConference{MeerKAT Science: On the Pathway to the SKA,\\
	25-27 May, 2016,\\
		Stellenbosch, South Africa}

\begin{document}
\section{Introduction}
The Zone of Avoidance (ZOA) remains an enigma in the study of large-scale structure, cosmic flow fields and the motion of the Local Group with respect to the Cosmic Microwave Background (e.g. \cite{ZOAR00,Loeb08}). Significant progress has been made in the whole-sky mapping of the large-scale galaxy distribution in the nearby Universe ($cz \la 15\,000$\,\kms\ \cite{Jarrett04, Huchra12}) to describe the cosmic web. However, due to dust obscuration and high stellar density these whole-sky surveys all 'avoid' a broad band of $\pm 5\deg$ to $\pm10\deg$ around the Galactic equator. This does not hold for systematic \HI\ surveys (e.g. \cite{HIPASS, HIZOA}), but so far these have been quite shallow. 

Local galaxy overdensities exert gravitational perturbations on the smoothly expanding Universe, resulting in the so-called peculiar velocity or cosmic flow fields. Our own Local Group (LG) partakes in such a flow, as evidenced by the dipole observed in the Cosmic Microwave Background (622~\kms\ \cite{Fixsen96}. Such flows are exceedingly well described through the analysis of peculiar velocity data \cite{CosFlow2, Scrimgeour16, Springob16}. However, the paucity of ZOA redshift data is endemic to all peculiar velocity surveys, and is a major limitation for our understanding of bulk flows. 

The direction of these flows, and the volumes over which they are coherent, remain controversial \cite{Hudson04, Erdogdu06, Kocevski06, Bilicki11}. Later results suggest a considerable mass excess just outside the boundaries and volumes of current systematic whole-sky redshift and peculiar velocity surveys ($cz \la 10\,000 - 16\,000$\,\kms), suspiciously close to the ZOA ($\ell: 270\deg - 330\deg$) \cite{Nusser11, Turnbull12, Springob14, Carrick15, CosFlow2, Scrimgeour16}. More recently the 6dFGS (6dF Galaxy Survey) and 2MTS (2MASS Tully-Fisher Survey) peculiar velocity data \cite{Springob14, Scrimgeour16, Springob16} found a residual bulk flow velocity of 273\kms\ that must have its origin due to a mass overdensity in the ZOA region close to Vela ($\ell \sim 270\deg$), arising from beyond 16\,000\,\kms.

This is exactly at the location in redshift space  where a major supercluster, dubbed the Vela Supercluster (VSCL), was recently discovered \cite{VSCLMN}. Although still sparsely sampled, it is enormous in extent ($\sim$\,115$\,\times\,$90~$h_{70}$\,Mpc), and has a significant influence on the motion of the Local Group (50\,\kms). Given its location on the sky it may well play an important role in explaining the above mentioned bulk flow results, and help reduce the misalignment angle between the CMB dipole and the motion of the LG.

The results and analysis of the VSCL to date is based on a sparsely-sampled spectroscopic survey over the outer edges of the ZOA ($|b| \la 5\deg$) where optical observations allow  good signal-to-noise spectra for the partially obscured galaxies. But no observational data exists as yet for VSCL at the lowest latitudes. Following a brief overview of the main properties of the Vela supercluster \cite{VSCLMN} in Sect.~2, we will present in Sect.~3 a MeerKAT observing strategy in early science mode to map the gas-rich spiral galaxy population of VSCL across the opaque part of the Milky Way, compare the predictions to simulations in Sect.4, and conclude with a summary in Sect.~5.

While the science goal of the VSCL MeerKAT survey project is self-contained, this project is a collaboration with the PI's and other members of the MeerKAT \HI\ Large Survey Projects (LSP) Laduma \cite{Laduma} and Fornax \cite{Fornax}. The hands-on experience with these early science \HI-data will yield invaluable experience towards the final preparation of the approved LSP's, in addition to benefiting from ample KAT-7 observing experience by various members of our team.
 
\section{The Vela Supercluster (VSCL)}

\subsection{Discovery and first results}

Details of the characteristics known so far about the Vela Supercluster are detailed in the discovery paper by Kraan-Korteweg et al. 2016 \cite{VSCLMN}. The results are based on a spectroscopic survey of partially obscured galaxies in the Vela survey region ($\ell = 272\fdg5 \pm 20\deg, b = 0\deg \pm 10\deg$). These included (i) the multi-object spectrometer (MOS) on the Southern African Large Telescope (SALT) \cite{SALT2015}, which are ideal for rich clusters at the suspected VSCL distance,  and (ii) the 2dF+AAOmega spectrograph on the 3.9m telescope of the Australian Astronomical Observatory (AAO), a perfect survey instrument with its 2-degree field and 392 fibres. It should be emphasised that the survey footprint is not contiguous, this is by design to sample as wide an area as possible in the enormous Vela region. It is dominated by the 25 AAOmega fields observed in 2014, plus earlier observed (unpublished) 6dF and Optopus ZOA data, but has little to no data below Galactic latitudes of $b < \pm 5\deg$. Combining the AAOmega and SALT redshift data with earlier unpublished data, led to a total of 4432 new redshifts.

\begin{figure}
\begin{center}
	\includegraphics[width=0.8\columnwidth]{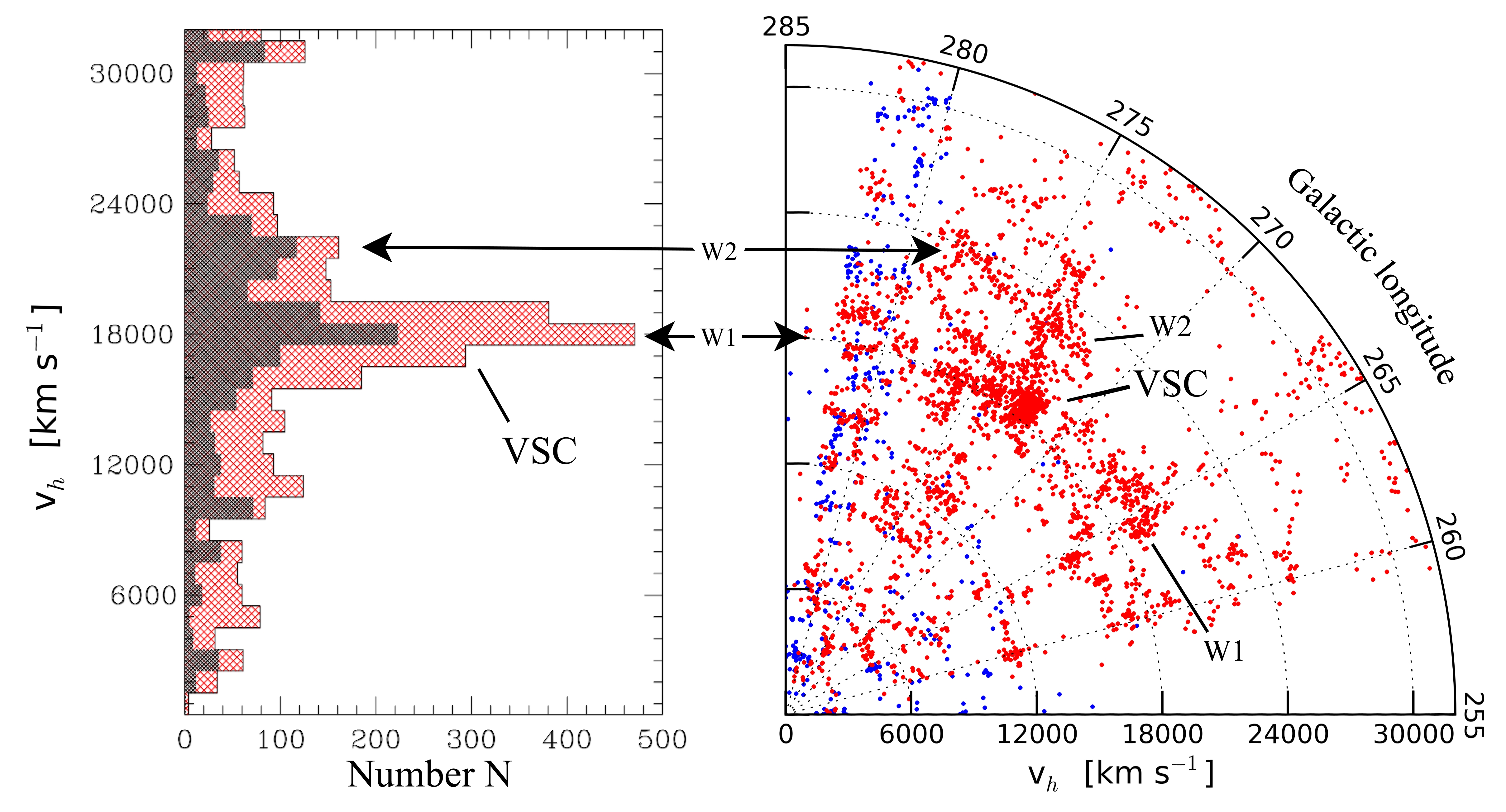}
    \caption{Velocity distribution of ZOA galaxies in the Vela Supercluster out to 32,000\kms, delimited by $(\ell,b) = (285\deg - 255\deg; \pm 10\deg)$. The left panel shows a histogram of all galaxies, grey-shaded are galaxies below the Plane. The right panel presents the redshift slice of same data. Red dots mark galaxies observed in our survey, blue dots redshifts in the literature (there is some overlap). The main two Walls of the VSCL are indicated.
    }
    \label{fig:wedge}
    \end{center}
\end{figure}

The VSCL is prominent over a surprisingly wide area on the sky ($30 \times 20\deg$) given its distance. More astonishing was its prominence.  Figure~\ref{fig:wedge} shows a radial velocity histogram (left panel) and a redshift wedge (right panel) along the Vela ZOA survey region $\ell,b = 285\deg - 255\deg; \pm 10\deg$. 

The velocity histogram shows a highly significant peak centered at $17\,000 - 19\,000$\kms, with broad shoulders ranging from $15\,000 - 23\,000$\kms, typical of superclusters. Indeed, its velocity distribution is nearly indistinguishable from the one based on the systematic mapping of the Shapley supercluster (SSC) \cite{Proust06}, one of the most massive superclusters in the nearby Universe. The grey-shaded area marks galaxies below the plane, to highlight the similarity of the velocity distribution on either side of the Plane, supporting continuity across the Plane.  

The right panel displays the distribution of the same galaxies.  The VSCL appears to consist of a primary wall structure (W1) at $\sim$\,18\,000\,\kms\ that can be traced over most of the length of the wedge (some of the gaps in the wall are due to gaps in survey coverage -- see also Fig.~\ref{fig:survey}). There is a second wall (W2) at slightly higher distances. The main wall W1 is prominent on both sides of the Galactic plane; the slightly more distant wall is visible only below the Plane. Taking account of the lack of data in wall W2 for  $\ell\sim 267\deg -272\deg$, it seems likely that W2 merges with W1 around $\ell\sim265\deg$. Numerous galaxy clusters ($\sigma > 400$\kms) are embedded within the walls.
Only a few are evident as X-ray clusters. However, the analysis in \cite{VSCLMN} highlights that less than one X-ray cluster would be expected in the ROSAT All Sky Survey given the constraints due to the Galactic foreground gas (which prevents detection at high gas column densities) as well as the Vela SNR for a uniform density. Otherwise, the morphology of the structure is fully consistent with a large, possibly assembling, supercluster (see e.g. \cite{Einasto14}). 

The most interesting question now is: How does VSCL compare to the well-studied and supermassive SSC? Will it affect the motion of the Local Group? Can VSCL help explain the residual bulk flow results? A first attempt to derive a qualitative estimate of the overdensity -- although limited to the area where the data set is not too affected by foreground obscuration: $|b| \ga 6\deg - 10\deg$ -- indeed confirms a substantial enhancement in galaxy counts (by a factor of $f = 1.2$ per square degree and magnitude bin), as well as a mass overdensity in a volume shell around the VSCL ($\delta \sim 0.6$) based on photometric redshifts which predicts a contribution to the LG motion of 50\,\kms. However, this analysis should be regarded as preliminary, because of the current non-uniformity of  redshift sampling and the lack of data within the inner part of the ZOA.

\subsection{Next steps in the exploration of VSCL}

Despite the over 4200 new spectroscopic redshifts in the wider Vela ZOA survey region \cite{VSCLMN, SALT2015}, the current spectroscopic coverage remains very sparse ($\sim15\%$ of the area in which the VSCL appears prominent: $\ell,b = 285\deg - 255\deg; \pm 10\deg)$, and was (is) feasible only for regions where the extinction is not too excessive ($A_{\rm B} \la 3\fm0$). As a next step towards arriving at a more uniform spectroscopic coverage, a plan is in place to use the Taipan instrument \cite{Taipan} in science verification mode in 2017 (led by M. Colless).  Taipan is a new spectrograph with an innovative starbugs optical fibre positioner on the 1.2m Schmidt telescope of the AAO, that can target up to $\sim$150 objects per $6\deg$ field.

However, optical spectroscopy can not penetrate the very low-latitude regions where optical extinction and/or star density are high. Here only systematic surveys in the 21cm neutral hydrogen line will prevail (see e.g. \cite{HIZOA}). Current radio telescopes are not sensitive or fast enough. But with the SKA Pathfinders now coming online, considerably deeper systematic \HI-surveys with increased spatial resolution are within our reach. As detailed in the next sections, MeerKAT is an ideal instrument to unveil the most obscured part of the VSCL. The final system temperature of MeerKAT ($T_{\rm sys} = 22$\,K), as measured on an actual MeerKAT dish (M063) are far superior compared to the original specifications, making MeerKAT one of the current fastest radio telescopes at L-band given its survey speed of $A_{\rm e}/T_{\rm sys} = 424$\,m$^2/$K \cite{Justin} and resolution of about 10\,arcsec for the full 64-dish array.
Our simulations have shown that a MeerKAT survey of VSCL can be realized within reasonable timescales with the interim array of 32 dishes, hence in early science commissioning or open time mode, and does not have to await the full 64-dish array. This makes a start in 2017 a feasible option.
 
\section{Observational requirements for MeerKAT observations of VSCL}
Our primary goal is to detect all galaxies at the VSCL distance with an \HI\ mass of $M_{\mathrm{HI}} \ge 3 \times 10^{9}$\,M$_{\odot}$, i.e. half an order of magnitude below the characteristic \HI-mass, log\,$M^*_{\rm HI} = 9.8$\,M$_{\odot}$ \cite{Zwaan05}. With the \HI-mass functions (HIMF) having their maximum \HI-mass volume density around log $M^*_{\rm HI}$, a statistically significant sample of gas-rich VSCL spiral galaxies can be expected. Spiral galaxies are excellent tracers of large-scale structures like walls \cite{Koribalski04}. In addition, we can use the resulting HIMF above our \HI-mass completeness limit, to determine the overdensity within the VSCL volume by scaling the \HI\ masses to a well-calibrated \HI-mass function. The so determined overdensity ought to be free of any dust obscuration effects that plague optical and even near-infrared surveys \cite{2MASX, KKTJ05}, except possibly in areas of extreme continuum contamination \cite{HIZOA}. The so derived overdensity can then be compared to the galaxy count enhancement and volume overdensity derived from extinction-corrected $K\deg$-band counts, respectively photometric redshift shells \cite{VSCLMN}, which find evidence for a significant galaxy and mass excess.

To calculate the sensitivity required to achieve this goal, we adopt a mean distance to VSCL of 260~$h_{\rm 70}$\,Mpc, which corresponds to the main peak in the radial velocity distribution of the VSCL overdensity ($cz \sim 18~000$\,\kms). The equivalent integrated flux density of a log\,$M_{\rm HI} = 9.5$\,M$_{\odot}$ galaxy at that distance is $F_\mathrm{HI}\sim~0.2$\,Jy\,\kms.  Imposing the requirement that $F_\mathrm{HI}$ be at a 5-$\sigma$ level within a 200\,\kms\ line width, and assuming a channel width of 10\,\kms, the required noise level is $\sigma \sim 1$\,mJy/beam per channel. If we use the above mentioned MeerKAT survey speed, but for the interim array of 32 dishes (M32), and enter that into the radiometer equation, an integration time of 30 minutes per pointing leads to a noise level of 1.03\,mJy/beam -- which obviously would reduce by a quarter for M64.  

\begin{figure}
\begin{center}
	\includegraphics[width=0.7\columnwidth]{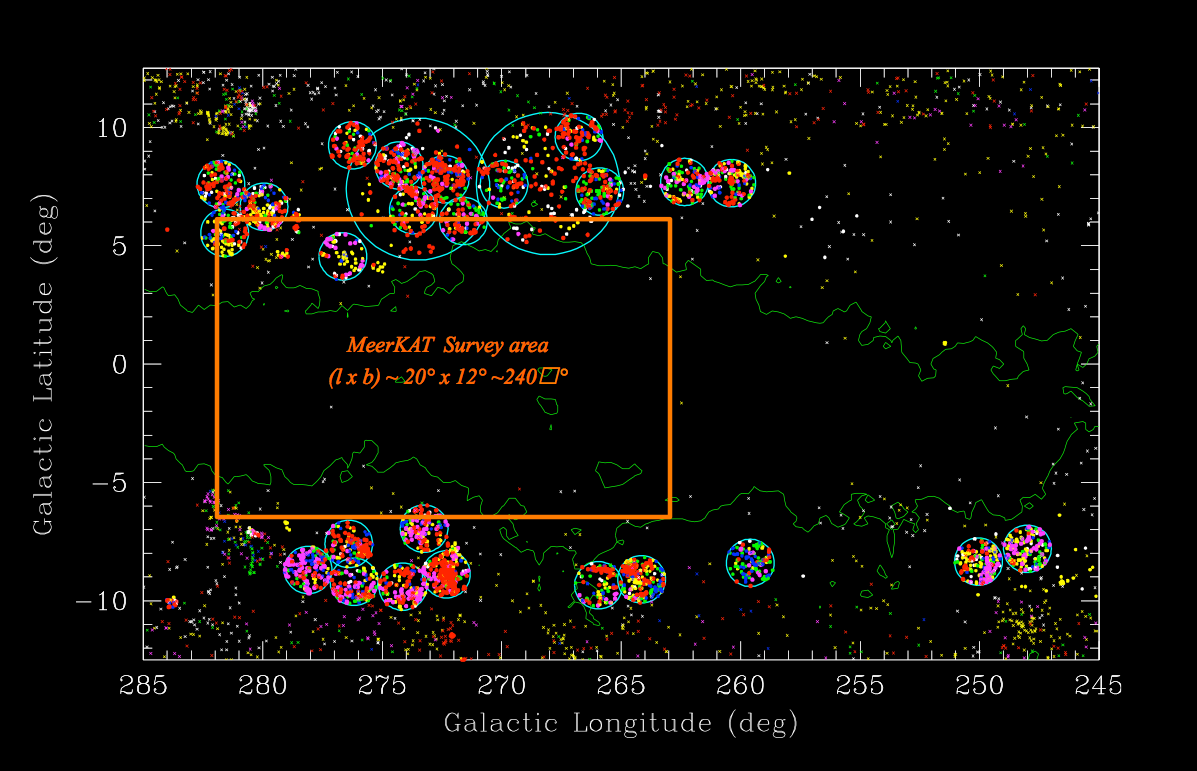}
    \caption{The envisioned MeerKAT survey region is overplotted on the Vela spectroscopic survey region in Galactic coordinates. The well-sampled two 6dF-fields and the 25 $2\deg$-AAOmega fields are indicated as cyan circles. Redshifts are colour-coded (white: $<$8000; yellow: 8--16000; red: 16--20000; magenta: 20--24000; blue: 24--32000\,\kms). The large dots mark new redshifts, crosses are previous redshifts. Contours indicate extinctions of $A_{\rm B} = 3\fm0$, above which the galaxy sample is incomplete, and spectroscopy nigh impossible.}
    \label{fig:survey}
    \end{center}
\end{figure}

Pursuing this as a pilot project -- and as a preparatory data set for the Fornax and Laduma MeerKAT surveys -- we restrict ourselves here to the minimal area that will lead to a scientifically insightful map of the extent and morphology of the VSCL walls that are hidden behind the highest Galactic dust column densities ($A_{\rm B} = 3\fm0$; see Fig.~\ref{fig:survey}). The largest concentration of VSCL galaxies on either side of the Galactic Plane are encountered around $\ell \approx 272\deg$. Figure~1 implies that both VSCL walls W1 and W2 pass through the so defined MeerKAT survey region. It might encompass a merger or an intersection of these two walls at lowest latitudes. This is also suggested in maps of the large-scale structures close to the Galactic Plane based on photometric redshift maps \cite{2MPZ, VSCLMN}. If the latter is correct, it is highly probable to find a massive cluster at this intersection. Rich clusters are normally found at the nodes of great-wall crossings, as evident in the cosmic web as well simulations of structure formation \cite{Springel05}. 

To optimally capture this complex structure, we hence propose a pilot survey area of about $20\deg \times 14\deg$ centered on $(\ell,b) = (272\deg, 0\deg$). The reason for surveying up to the latitudes of $b = \pm 7\deg$ rather rather than $b \le \pm 5\deg$ is the resulting overlap with the existing spectroscopic data set. This allows us to tie the 'blind' \HI-survey data to the spectroscopically surveyed region. With the large fraction of emission line galaxies identified in our spectroscopic sample, the \HI\ detection rate is expected to be significant.

The survey region is centered on the Milky Way, where continuum sources are rife. We therefore prefer to Nyquist-sample the hexagonal grid of MeerKAT pointings used to cover the target sky area. This assures uniform sensitivity over the survey area, allowing a firm derivation of the completeness limit and hence \HI-mass overdensity. The first holographic measurement of the MeerKAT dishes yield FWHM of $1.15 (\lambda / D)$. This translates into a FWHM$ = 1\fdg02$ at the zero-redshift wavelength of the 21cm line. A Nyquist sampling (spacing $= \mathrm{FWHM} / \sqrt{3}$) then translates into pointing offsets of $\sim 0\fdg6$. Therefore, a grid of $20\deg \times 14\deg$ would require 800 pointings. But with the Nyquist sampling every patch on the sky gets double the integration time compared to an individual pointing. The total observing per pointing can be therefore be reduced to 15\,min per pointing to reach the anticipated sensitivity. The total observing time for the outlined survey region with M32 then adds up to $\sim 200$\,hrs (or 50 hrs with the full M64 array), not including overhead.

\section{Comparison with simulations}
The next question is: how well will we be able to trace the extent of the supercluster and its wall with the envisioned observational set-up? To test this we have run some simulations. We used the catalogue of evaluated galaxy properties from Obreschkow \& Meyer \cite{Obresch14} to glean some insight into the possible sky distribution of the galaxies to be detected with the outlined MeerKAT survey strategy.  The catalogue spans a sky area of 100 deg$^2$ for the redshift range $z=0-1.2$.  For many galaxies it presents detailed \HI\ properties as well as auxillary optical properties. The catalogue is based on the SKA Simulated Skies semi-analytic simulations, and therefore on the physical models described in a series of papers led by Obreschkow \cite{Obresch09a, Obresch09b, Obresch09c}.

\begin{figure}
\begin{center}
	\includegraphics[width=0.9\columnwidth]{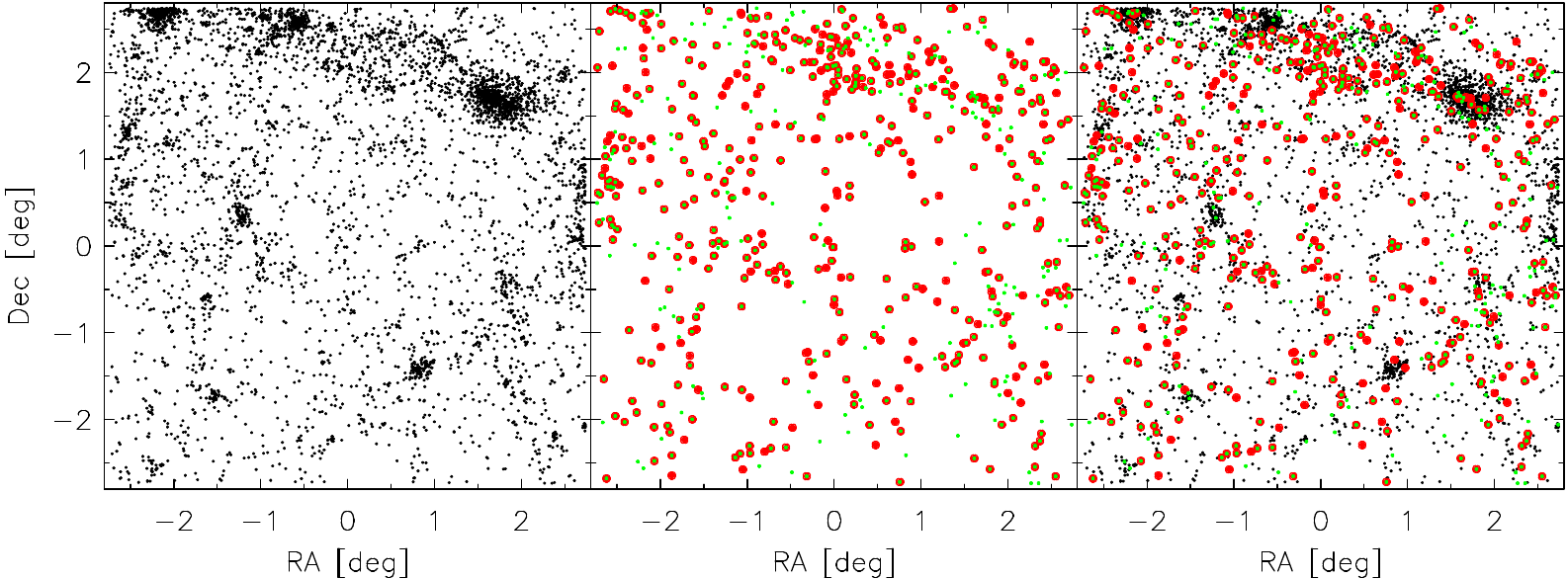}
    \caption{Sky distribution of galaxies from the Obreschkow~\&~Meyer catalogue, spanning the redshift range $z=0.048 - 0.073$.  Left panel: all 5072 galaxies in the volume.  Centre panel: galaxies that have an average flux density greater than 5$\sigma$ in a channel of frequency width $df=47$~kHz (filled red circles) and galaxies more massive than log $M_{\mathrm{HI}} \ge 9.5$\,M$_{\odot}$ (smaller filled green circles).  Right panel: all data from left and centre panels shown together.}
    \label{fig:sims}
    \end{center}
\end{figure}

From the catalogue we extract the properties of all galaxies spanning a sky area of 30 deg$^2$ within the redshift range $z=0.048 - 0.073$ ($cz \sim 15\,000 - 22\,500$\,\kms), i.e. the full velocity range within which the Vela supercluster raises its head (see Fig.~\ref{fig:wedge}).  The left panel of Fig.~\ref{fig:sims} shows the distribution of all 5072 galaxies (including early type galaxies) in that volume range. Interestingly, this simulation reveals a wall-like structure with a massive cluster embedded right at the VCSL distance,  with some additional filamentary structures that host smaller clusters and groups.  The central panel now shows all the galaxies we expect to detect with MeerKAT32 for an integration of 15min per pointing for a Nyquist sampled grid. The small, green filled circles are the galaxies that are more \HI-rich than our survey completeness limit of log $M_{\mathrm{HI}}=9.5$\,M$_{\odot}$ -- 118 objects over the 30 deg$^2$ of the simulation. The larger, red filled circles mark galaxies that have a flux density greater than 5-$\sigma$ for any observed line-width larger than 10 \kms. There are 465 such objects. Their $M_{\mathrm{HI}}$ reaches down to $\sim 8\times10^8$ M$_{\odot}$ although the sample is incomplete down to these levels. The panel on the right shows the red and green sub-samples together with the full sample of 5072 galaxies. The simulations clearly show, that we will recover the major walls very well through the detection of their more \HI-massive spirals. We also recover the thinner filaments very well. Not surprisingly, the detection rate is low in the highest density clusters, which is dominated by early type galaxies, and may also be affected by gas loss through ram pressure stripping or other gas removal processes caused by the dense environment. 

It is satisfactory to note, that we will not suffer from spatial confusion. The simulations reveal that the typical separation between \HI-massive galaxies peaks around 10\,arcmin, while there are virtually none that lie closer to each other than 60\,arcsec, hence much farther apart than the angular separation that we will obtain with either M32 or M64. (Of course, even sources relatively close on the sky will generally have a significantly different velocity along the line of sight at the proposed resolution of $\sim10$ \kms.)

The data were extracted from simulations within an area of 30 square degrees. The proposed VSCL MeerKAT survey area  will be a factor of $f = 9.3$ higher, hence also the predicted number of \HI\ detections. This therefore leads to a total of 4330 galaxies above the $5\sigma$-treshhold, and 1100 galaxies above the \HI\ mass completeness limit of log $M_{\mathrm{HI}} \ge 9.5$\,M$_{\odot}$. Note that this will constitute a lower limit 'if' this region is similarly overdense as the latitude strips above and below the MeerKAT survey region (a factor of $f = 1.2$ in galaxy counts, and a volume overdensity in a shell around VSCL of $\delta \sim 0.6$ \cite{VSCLMN}). In either case, the number of \HI\ detections will be more than sufficient to determine an overdensity based on the HIMF.

While the survey strategy is optimised for uncovering the obscured part of the VSCL, the observations will use the full L-band frequency range of the receiver. This will allow ample testing for other planned \HI\ MeerKAT surveys, like Fornax and Laduma:\\
\noindent{\sl -- Fornax:} The number of detections at the lower redshifts is substantial, including a large number of low-$M_{\mathrm{HI}}$ dwarfs. For instance, looking at the 5-$\sigma$ detection rate of galaxies within a volume delimited by $cz < 5000$\,\kms, we should uncover around 1200 galaxies above an \HI-mass completion limit of $M_{\mathrm{HI}} \ge 10^8$~M$_{\odot}$, the majority of them dwarfish (below $\la 10^9$~M$_{\odot}$). The pipeline (mosaic) testing and exploration of this data set will prove of interest to the Fornax team \cite{Fornax}. \\
\noindent{\sl -- Laduma:} The simulations reveal that the number of HI-massive galaxies, generally defined as galaxies with \HI-masses above log $M_{\mathrm{HI}} \ge 10.0$\,M$_{\odot}$ (e.g. \cite{Catinella15}), will yield around 3000 galaxies up to the redshifts of $z < 0.2$, with a further 1000 \HI-massive galaxies between the redshift range $0.2 < z < 0.5$ -- if such massive \HI-galaxies do exist in equally high numbers in the earlier Universe. Hence a highly interesting test bed towards the Laduma science goals of galaxy and HIMF evolution, and a challenging test of the available software tools for extracting galaxies at these high redshifts from this pilot MeerKAT data set.

\section{Summary}

The recently unveiled Vela supercluster \cite{VSCLMN} belongs amongst the more massive and extended superclusters  in the nearby Universe. Its overdensity is evident over a large area on the sky ($25\deg \times 20\deg$, respectively $\sim$\,115$\,\times\,$90~$h_{70}$~Mpc). Calculations find that its contribution to the motion of the LG might be as large as 50\kms, decreasing the misalignment angle with the CMB dipole by about $\sim 20-25\%$. This analysis is based on a spectroscopic skeleton sampling over the originally suspected extent of the supercluster \cite{VSCLMN,SALT2015}, assuming continuity of structures across the innermost part of ZOA, where the thickness of the dust column densities prevent further optical spectroscopy. 

We have presented a MeerKAT survey scenario to map the VSCL across this obscured band and get a full census of the supercluster, its morphology and mass. We have shown that a Nyquist-sampled survey region covered by 800 pointings (15\,min per pointing) will yield the sensitivity to map the supercluster through the detection of its gas-rich spiral galaxy population, and an overdensity determination through HIMF fitting. The optimal survey region of $(\ell,b) \sim (272\deg\pm10\deg, 0\deg\pm7\deg$; 280 square degrees) could be covered with M32 in 200\,hrs. We highlight the use of the M32 array for this survey as an early science project given its high-impact cosmological implications on the one side, and its suitability as precursor for the planned MeerKAT \HI-surveys, in particular Fornax and Laduma, on the other hand.

\section*{Acknowledgements}
We thank  Laura Richter from the SA SKA Office for useful discussions, and the other members of the original VSCL discovery team (M. Bilicki, M. Colless, A. Elagali, H. Boehringer, G. Chon), and . All South African authors acknowledge the research support they received from the South African NRF; E.E., B.F. thank the SA SKA Office as well. This project has received funding from the European Research Council (ERC) under the European Union’s Horizon 2020 research and innovation programme (grant agreement No 679627). The VSCL supercluster redshift results are based on observations taken at the AAO, as well as data obtained with SALT. This publication makes use of data products from the Two Micron All Sky Survey, which is a joint project of the University of Massachusetts and the Infrared Processing and Analysis Center/California Institute of Technology, funded by the National Aeronautics and Space Administration and the National Science Foundation.

\end{document}